\documentclass[twocolumn,slac_two]{revtex4}
\usepackage{graphicx}
\usepackage{fancyhdr}
\newcommand{\fermi}{{\it Fermi} }

\begin{document}

\title{GBM Monitoring of Cyg X-1 During the Recent State Transition}


\author{G. L. Case, S. Baldridge, M. L. Cherry}
\affiliation{Department of Physics and Astronomy, Louisiana State University, Baton Rouge, LA 70803, USA}

\author{A. Camero-Arranz, M. Finger}
\affiliation{Universities Space Research Association, Huntsville, AL 35805, USA}

\author{P. Jenke, C. A. Wilson-Hodge}
\affiliation{NASA Marshall Space Flight Center, Huntsville, AL 35182, USA}

\author{V. Chaplin}
\affiliation{Department of Physics, University of Alabama, Huntsville, Huntsville, AL 35899, USA}

\begin{abstract}
Cygnus X-1 is a high-mass x-ray binary with a black hole compact object. It is normally extremely bright in hard x-rays and low energy gamma rays and resides in the canonical hard spectral state. Recently, however, Cyg X-1 made a transition to the canonical soft
state, with a rise in the soft x-ray flux and a decrease in the flux in the hard x-ray and low energy gamma-ray energy bands. We have been using the Gamma-Ray Burst Monitor on Fermi to monitor the fluxes of a number of sources in the 8--1000 keV energy range, including Cyg X-1. We present light curves of Cyg X-1 showing the flux decrease in hard x-ray and low energy gamma-ray energy bands during the state transition as well as the several long flares observed in these higher energies during the soft state. We also present preliminary spectra from GBM for the pre-transition state, showing the spectral evolution to the soft state, and the post-transition state.
\end{abstract}

\maketitle

\thispagestyle{fancy}


\section{Introduction}
Cygnus X-1 is a high-mass x-ray binary system and was the first system in which the compact object was thought to be a black hole. It is a microquasar, as confimed by the detection of relativistic jets \citep{stirling01}, and displays variability on time scales from minutes to months. Cyg X-1 spends most of its time in the so-called hard state, characterized spectrally by a hard power law with an exponential cutoff at $>100$ keV.  The hard state can also be modeled by thermal Componization of cool seed photons in a hot electron plasma with $kT \geq 40$ keV.  The soft x-ray blackbody emission is weak in this state.  

Occasionally, Cyg X-1 transitions into a thermal dominant, or soft, state.  This state is characterized by a strong thermal component arising from a geometrically thin but optically thick accretion disk.  The hard x-ray component is very weak and spectrally steep in the soft state.  Cyg X-1 does not remain in the soft state for long (typically $<100$ days) before returning to the hard state.  Additionally, there are other black hole spectral states, such as the intermediate state and steep power law state (see ref \citep{remillard03}), in which Cyg X-1 has been observed to reside.  

Monitoring Cyg X-1 in hard x-rays, particularly above 100 keV, is important as it allows contraints to be made on the spectral cutoff energy and Comptonization temperature as the system transitions from the hard to the soft state.  These measurements can then help constrain models of the physical emission mechanisms, including the contribution of the jets.

The Gamma-ray Burst Monitor (GBM) \citep{meegan09} is the secondary instrument onboard the \fermi satellite. It consists of 12 NaI detectors $5^{\prime\prime}$ in diameter by $0.5^{\prime\prime}$ thick mounted on the corners of the spacecraft and oriented such that they view the entire sky not occulted by the Earth. GBM also contains 2 BGO detectors $5^{\prime\prime}$ in diameter by $5^{\prime\prime}$ thick located on opposite sides of the spacecraft. None of the GBM detectors have direct imaging capability.

Known sources of gamma-ray emission can be monitored with non-imaging detectors using the Earth occultation technique, as was successfully demonstrated with BATSE \citep{ling00,harmon02}. When a source of gamma rays is occulted by the Earth, the count rate measured by the detector will drop, producing a step-like feature. When the source reappears from behind the Earth's limb, the count rate will increase, producing another step. These steps are fit to a model incorporating the atmospheric attenuation and the changing instrument response, along with a quadratic background, to determine the detector count rate due to a particular source. Up to 31 steps are possible for a given source in a day, and these steps are summed to get a single daily average flux.  The Earth occultation technique requires an input catalog of sources and their positions, and these sources can then be monitored by GBM.  The technique and initial results are given in references \citep{wilson09} and \citep{case11}.

The light curves are generated using the GBM CTIME data, with its 8 broad energy channels. The spectra are produced using the GBM CSPEC data, with its 128 energy channels, binned up into 16 energy bands from 10--400 keV. Because our occultation fitting code currently is limited to 8 energy channels at a time, the 16 bins are broken into 2 pieces with separate PHA files generated for the low energy and high energy sections. Both the low energy and high energy data sets are then fit jointly in XSPEC.

\section{Light Curves}
\begin{figure}[tb]
\centering
\includegraphics[width=84mm]{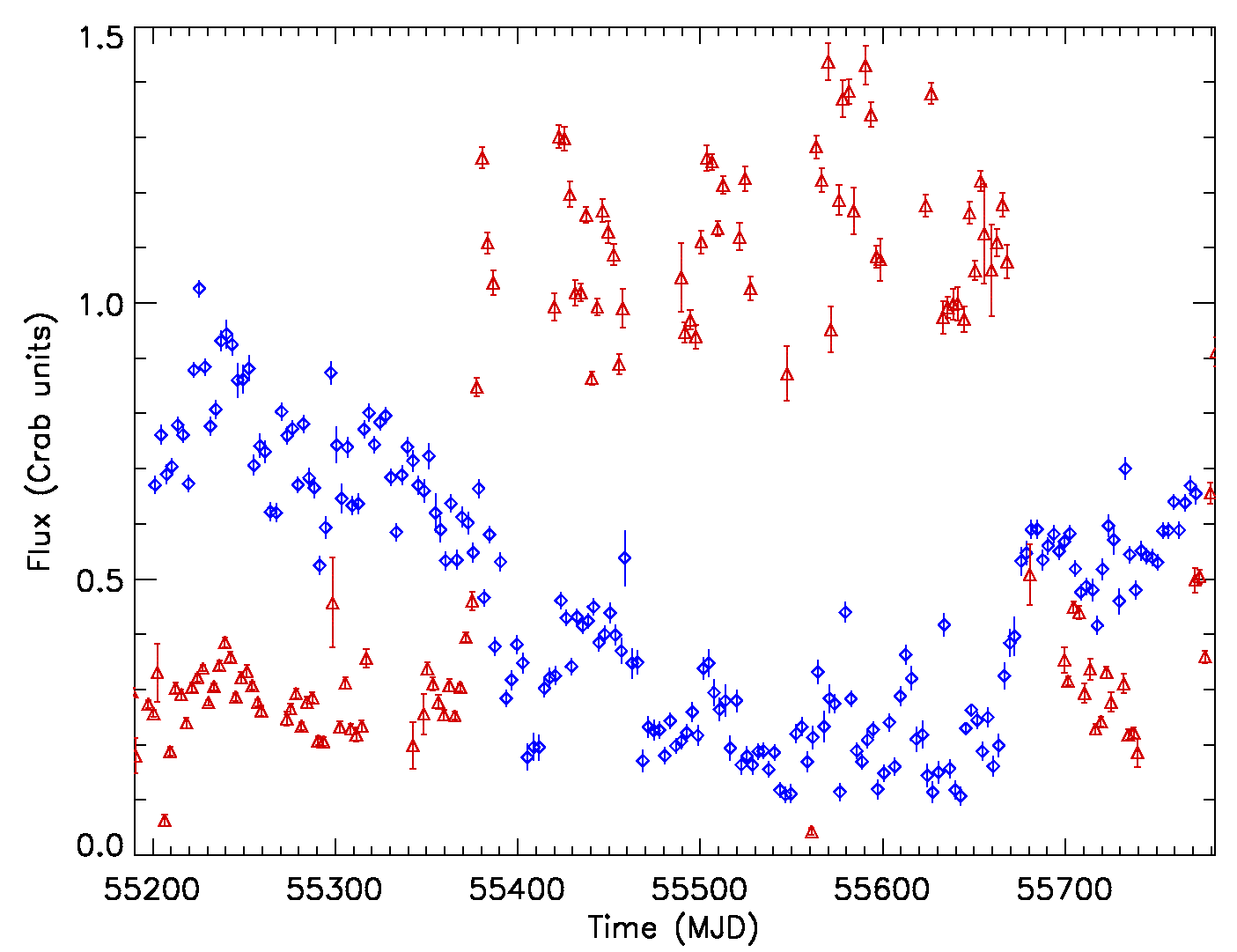}
\caption{GBM (blue diamonds) 12--50 keV and MAXI/GSC (red triangles) 1.5--4.0 keV light curves showing the typical anti-correlation between the soft and hard x-rays
as the system transitioned from the hard state to the soft state and back again. The data are binned 3 days per data point.} \label{maxi_lc}
\end{figure}

Cygnus X-1 has been monitored by the GBM Earth Occultation Team since \fermi science operations began on 2008 August 12 (MJD 54690).  During this time, Cyg X-1 was in the canonical hard state until a hard-to-soft state transition began at the beginning of 2010 July.  On MJD 55374, MAXI detected a rapid rise in the soft 2--4 keV band \citep{negoro10}, which combined with the decrease in the low energy gamma-ray flux \citep{wilson10} indicated a transition to a thermally-dominated soft state. Figure \ref{maxi_lc} shows the MAXI/GSC 1.5--4 keV light curve, binned 3 days per data point, overlaid on the GBM 12--50 keV broad band light curve showing the typical anti-correlation between the soft and hard x-rays.

\begin{figure}[t]
\centering
\includegraphics[width=85mm]{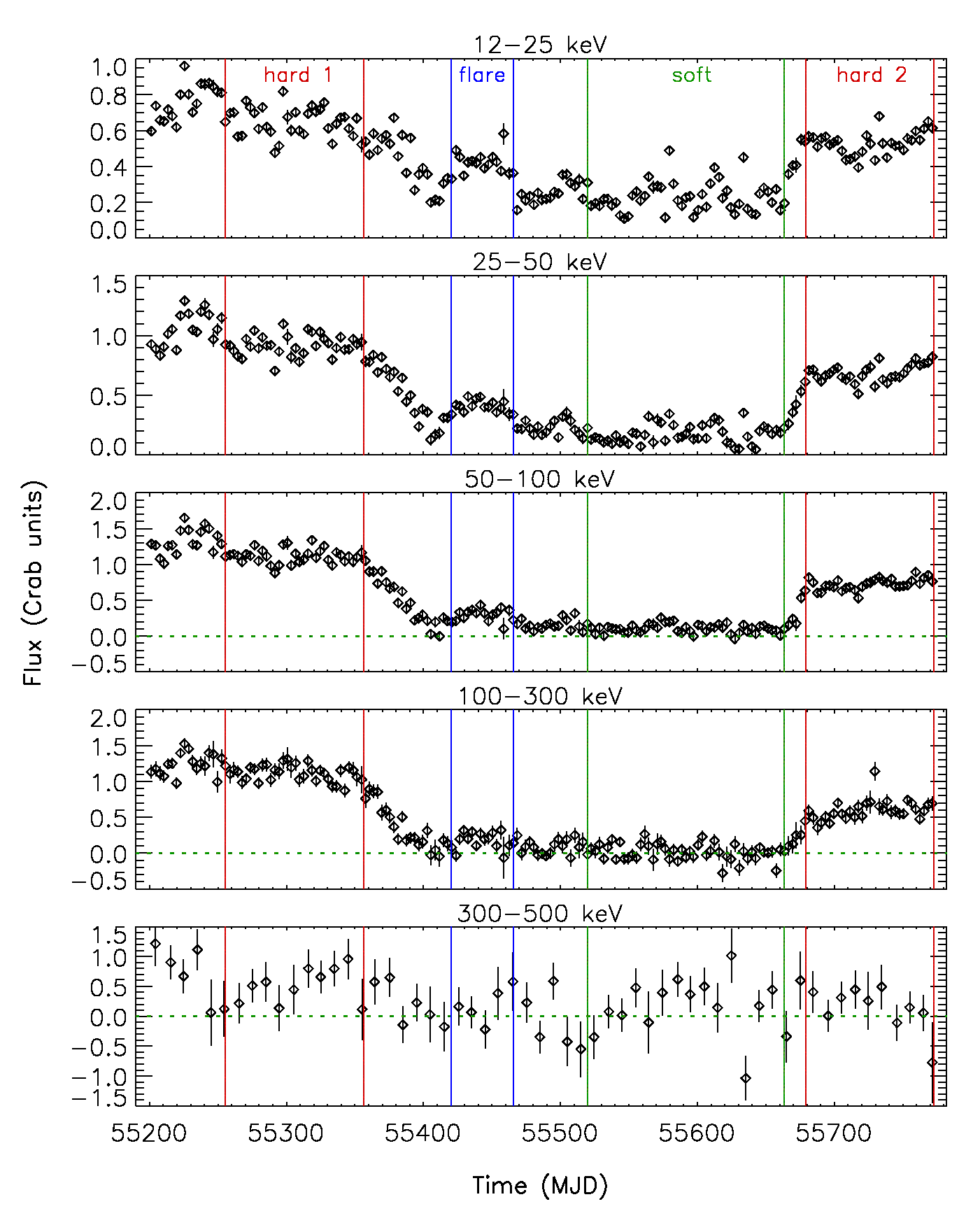}
\caption{GBM light curves in 5 of the broad CTIME energy bands for Cyg X-1. The data points represent 3-day averages for the 12--25, 25--50, 50--100, and 100--300 keV bands, and 10-day averages for the 300--500 keV band.  The regions from which spectra were obtained are marked by the veritcal lines (see Figures 3-5).} \label{gbm_lc}
\end{figure}

Figure \ref{gbm_lc} shows the light curves for GBM occultation data in 5 broad energy bands from 12--500 keV starting about 150 days before the hard-to-soft transition.  The hard-to-soft transition in the 100--300 keV band appears to begin about 20 days earlier than in the 12--25 keV band. The initial decline is much steeper in the 100--300 keV band, dropping by a factor of $\sim8$ from 1100 mCrab on MJD 55355 to less than 200 mCrab by MJD 54000, while the decline in the 12--25 keV band is about a factor of $\sim2$, from about 650 mCrab on MJD 55378 to about 300 mCrab on MJD 55405.  

Note that there is significant, though variable, emission above 300 keV up until about MJD 55372.  After the hard-to-soft transition, the observations above 300 keV were consistent with no detection up until at least MJD 55550. From MJD 55553 to 55659, the flux detected in the 300--500 keV band is marginally significant at $348\pm104$ mCrab.  After the transition back to the hard state, the measurements in the 300--500 keV band are consistent with a non-detection.

After the initial decline, Cyg X-1 rebrightened as part of a broad ``flare'' lasting $\sim45$ days.  The 12--25 keV flux peaked at about 500 mCrab before dropping back to about 200 mCrab.  This broad flare was seen in the 100-300 keV band as well, though it peaked at less than 400 mCrab before dropping to undetectable levels.  Several other broad flares are evident in the lower energy bands while Cyg X-1 was in the soft state.

Cyg X-1 remained in the soft state for nearly 300 days before making the transition back to the hard state in the middle of 2011 April \citep{grinberg11}.  Unfortunately, Cyg X-1 was not visible to MAXI during the beginning of this transition, and once Cyg X-1 became visible again, the 1.5--4 keV flux had nearly reached its canonical hard state level.

Once Cyg X-1 completed the transition back to the hard state (hereafter referred to as hard 2), it did not recover fully to the hard state flux levels before the state changes (hereafter referred to as hard 1).  The flux in the 12--25 keV band reached $\sim80\%$ of its pre-transition hard 1 level, while 100--300 keV flux only reached $\sim50\%$ of the pre-transition hard 1 level. 
This indicates that while Cyg X-1 was back in the hard state, the hard 2 spectrum is softer than it was for the $\sim100$ days before the hard-to-soft transition.  

\section{Spectra}
\begin{figure}[tb]
\centering
\includegraphics[width=94mm]{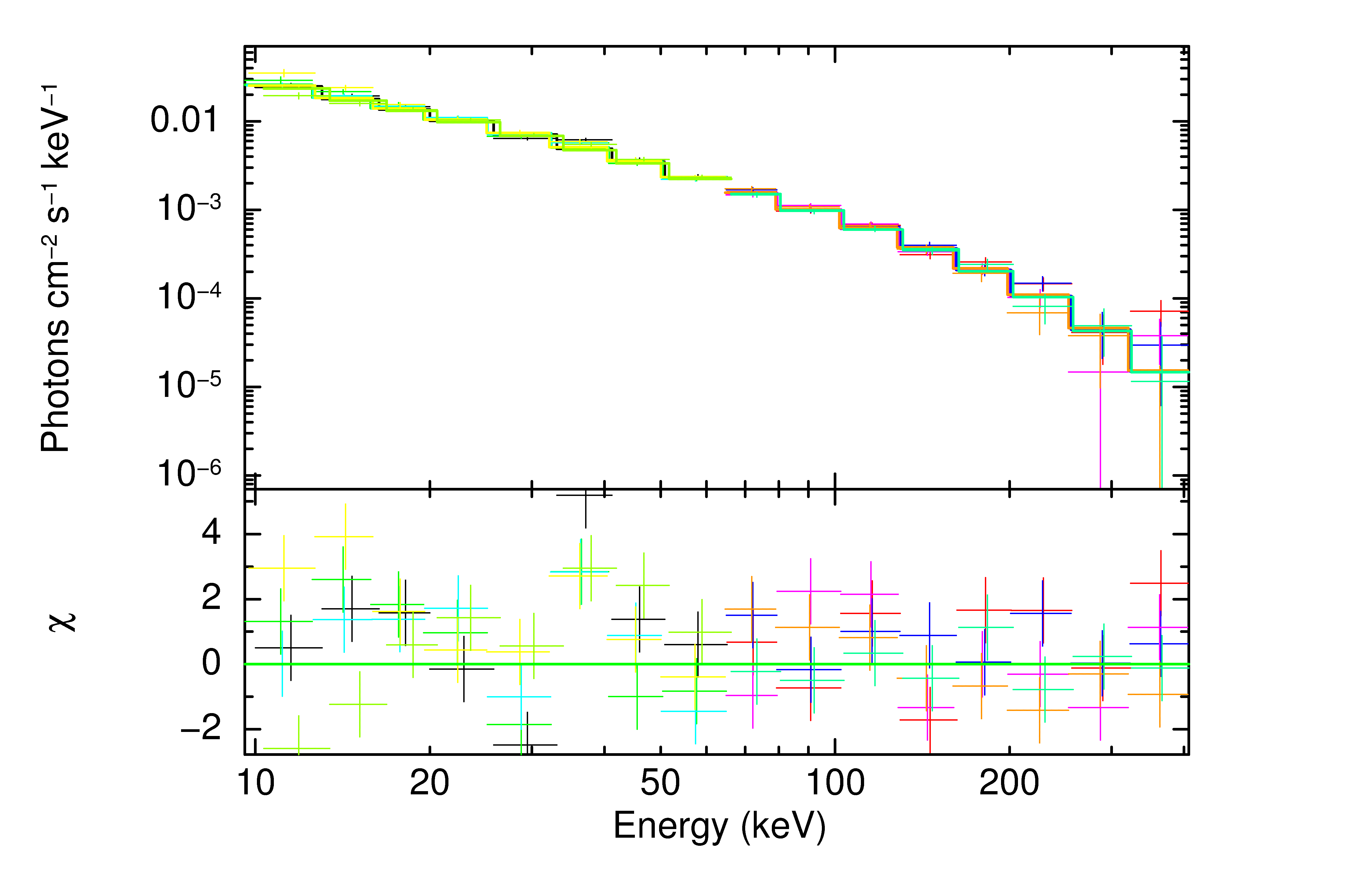}
\caption{GBM spectrum of Cyg X-1 in the pre-transition (hard 1) hard state.} \label{sp_hard1}
\end{figure}

The GBM CSPEC data have been used to generate the count rates for Cyg X-1 in 16 energy bins from 10--400 keV, from which preliminary spectra have been extracted. The average spectrum for the hard 1 state is obtained from the sum of the data from MJD 55255--55356 and is plotted in Fig.~\ref{sp_hard1}. Only the five detectors with the largest exposure and detector angles $<60^{\circ}$ are plotted. Following reference \citep{bel06}, the data are fit to an absorbed Comptonization model with reflection (\texttt{wabs (reflect * compTT)} in XSPEC notation). As in reference \citep{bel06}, the column density is frozen at $N_H = 0.6 \times 10^{22} {\rm cm}^{-2}$, $kT_0$ is frozen at 0.2 keV, and the inclination angle is fixed at $45^{\circ}$. The fit results are shown in Table \ref{sp_param}. The temperature of the Compton cloud and optical depth are slightly lower than, and the reflection component similar to, that reported by INTEGRAL for observations from 2002--2004 \citep{bel06}. Fitting the GBM data to an absorbed cutoff power law gives essentially the same $\chi^{2}_{\nu}$, with a hard power law index of $1.21 \pm 0.02$ and and a rather high cutoff energy of $113\pm5$ keV.

\begin{table}[t]
\begin{center}
\caption{Spectral parameters}
\begin{tabular}{|c|c|c|c|c|}
\hline                         & \textbf{Hard 1} & \textbf{Flare} & \textbf{Soft} & \textbf{Hard 2} \\
\hline $kT$(keV)               & $52.1\pm2.9$    &                &               & $40.7\pm3.5$  \\
\hline $\tau$                  & $1.37\pm0.07$   &                &               & $1.37\pm0.07$ \\
\hline $\Omega/2\pi$           & $0.35\pm0.04$   &                &               & $0.35\pm0.04$ \\
\hline $\chi^{2}_{\nu}(\nu)$   & 1.58(3580)      &                &               & 1.34(2860)     \\ 
\hline
\hline $\alpha$                & $1.21\pm0.02$   & $1.41\pm0.08$  & $1.80\pm0.11$ & $1.23\pm0.03$ \\
\hline $E_c$(keV)              & $113\pm5$       & $48\pm6$       & $56\pm11$     & $77\pm4$      \\
\hline $\chi^{2}_{\nu}(\nu)$   & 1.58(3581)      & 1.31(1389)     & 1.25(4509)    &  1.34(2861)     \\
\hline
\end{tabular}
\label{sp_param}
\end{center}
\end{table}

\begin{figure}[tb]
\centering
\includegraphics[width=86mm]{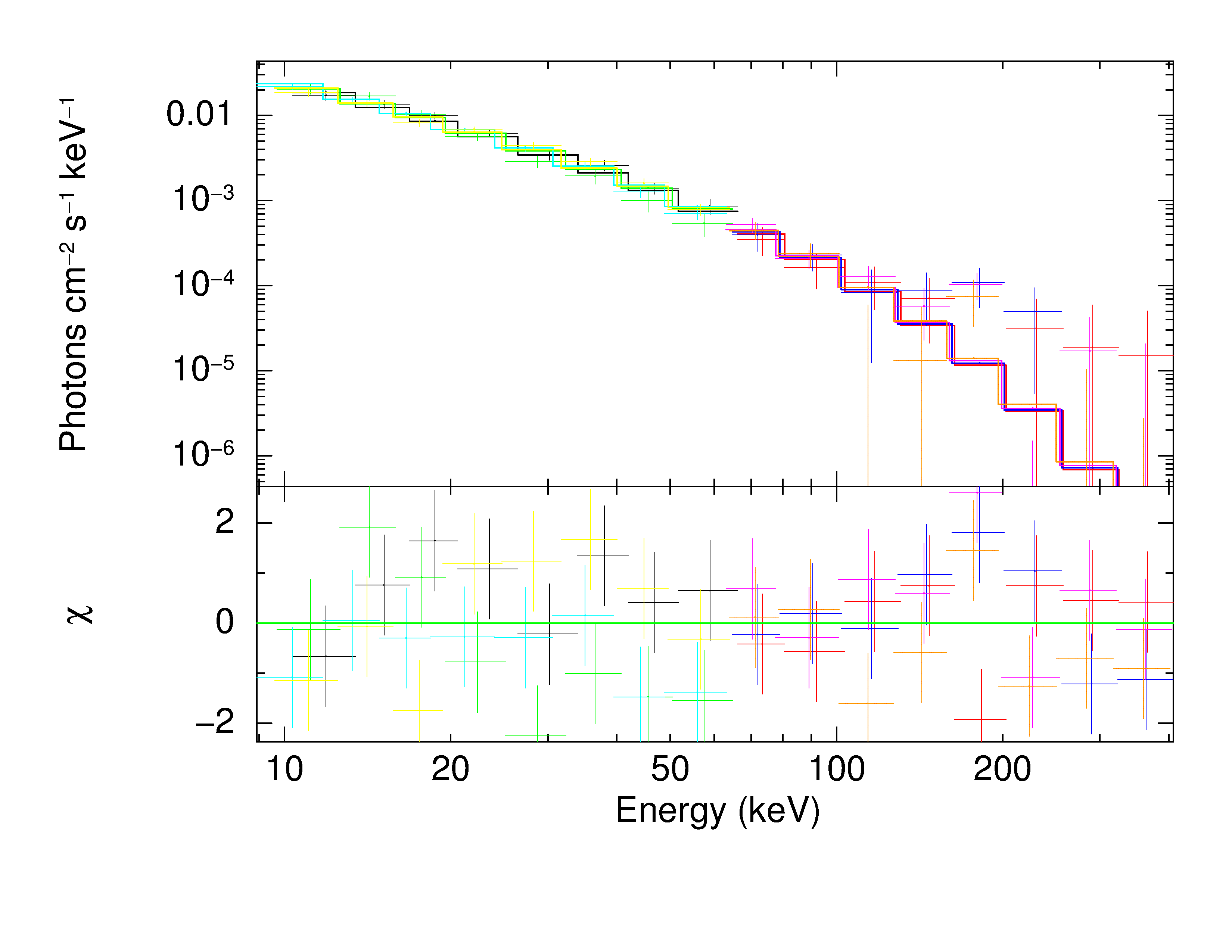}
\caption{GBM spectrum of Cyg X-1 in the flare state.} \label{sp_flare}
\end{figure}

\begin{figure}[tb]
\centering
\includegraphics[width=94mm]{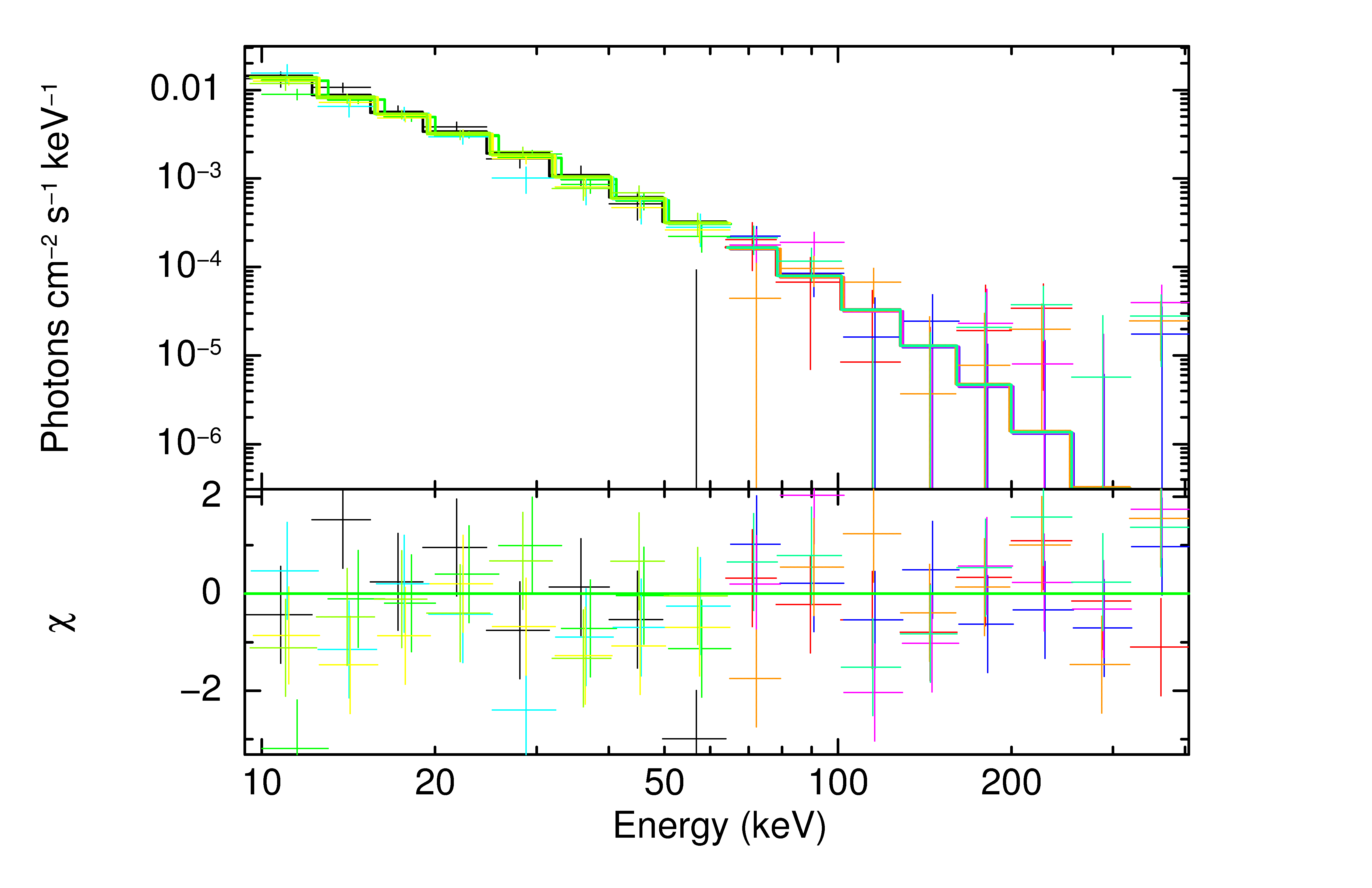}
\caption{GBM spectrum of Cyg X-1 in the soft state.} \label{sp_soft}
\end{figure}

The average spectrum in the soft state (MJD 55520--55663) is shown in Fig.~\ref{sp_soft}.  All five of the detectors that viewed Cyg X-1 with a detector angle $<40^{\circ}$ are shown.  There are not enough counts in the high energy bins to constrain the Comptonization model, so an absorbed cutoff power law model was used.  The fit results are shown in Table \ref{sp_param}.  As would be expected, the power law index is steeper and the cut off energy lower than in either of the hard 1 or hard 2 states.

After the initial decline in the light curve, there was a broad flare lasting $\sim45$ days. The average spectrum over the time period MJD 55420--55465 is shown in Fig.~\ref{sp_flare}. Only the four detectors with the largest exposure and detector angles $<40^{\circ}$ are shown. As with the soft state, the Comptonization model is not well constrained, so the spectrum is again fit to an absorbed cutoff power law model. As can be seen in Table \ref{sp_param}, the photon index is flatter than that of the soft state, while the cutoff energy is nearly the same as in the soft state, suggesting that the system was in a state intermediate between the hard and soft state.

\begin{figure}[tb]
\centering
\includegraphics[width=94mm]{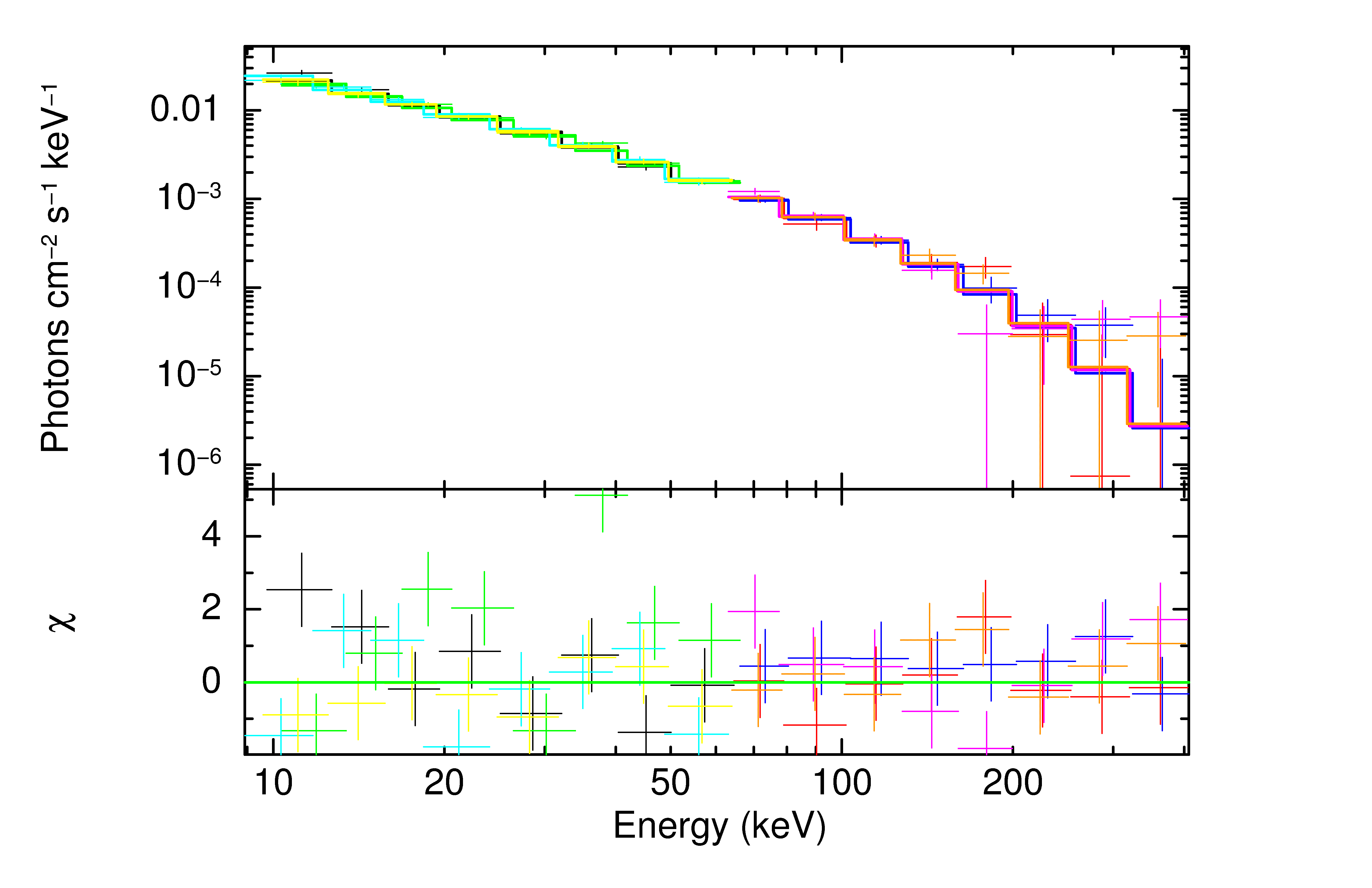}
\caption{GBM spectrum of Cyg X-1 in the post-transition (hard 2) hard state.} \label{sp_hard2}
\end{figure}

The average spectrum for the hard 2 state (MJD 55679-55772) is shown in Fig~\ref{sp_hard2}. Only the four detectors with the largest exposure and detector angles $<40^{\circ}$ are shown. Again following reference \citep{bel06}, the data are fit to an absorbed Comptonization model with reflection, with the fit values given in Table \ref{sp_param}. Fitting the data to an absorbed cutoff power law gives the same $\chi^{2}_{\nu}$, with a hard power law index of $1.23\pm0.03$, the same as in the pre-transition hard 1 state. However, the cutoff energy is lower (and the plasma temperature for the Comptonization model is lower) in the post-transition hard 2 state. This is consistent with the fact that Cyg X-1 was in an unusually hard state between 2006 and 2010 \citep{nowak11}.

\section{Conclusion}
GBM has been monitoring Cyg X-1 for the last $\sim3$ years using the Earth Occultation Technique. Cyg X-1 had been in the canonical hard state until July 2010, when it underwent a transition to the soft state. The GBM light curves suggest that the decline in the higher energies led the decline in the lower energies by several days, and that the declines were steeper in the higher energy bands, consistent with the expected softening of the spectrum.

Preliminary spectra have been generated using the GBM CSPEC data. The initial hard state spectrum (hard 1) is similar to that measured during previous Cyg X-1 hard states.  The spectrum around the peak of the first broad ``flare'' after the transition suggests that Cyg X-1 was in an intermediate state before dropping back into the soft state.  Cyg X-1 made the transition back to the hard state (hard 2) beginning in April 2011.  The spectra for the hard 2 state was softer than the hard 1 state. This is consistent with the fact that Cyg X-1 was in an unusually hard state between 2006 and 2010 \citep{nowak11}.  

More recently, Cyg X-1 has once again made the transition back to the soft state \citep{negoro11,grinberg11b,case11b}.  This is unusual, as this source has not made transitions to the soft state so close together in time in at least the last 20 years.  We will continue to monitor Cyg X-1, as well as other black hole candidates, for future outbursts and/or state transitions.

\bigskip % extra skip inserted
\begin{acknowledgments}

\end{acknowledgments}
This work is supported by the NASA Fermi Guest Investigator program. At LSU, additional support is provided by NASA/Louisiana Board of
Regents Cooperative Agreement NNX07AT62A.

\bigskip % extra skip inserted
%\bibliography{Case_Fermi_Proceedings}
%\begin{thebibliography}{9}   % Use for  1-9  references

\end{document}